\documentclass[runningheads]{llncs}
\usepackage[T1]{fontenc}
\usepackage{graphicx}
\usepackage{booktabs}
\usepackage{multirow}
\usepackage{amsmath}
\usepackage{amssymb}
\usepackage{amsfonts}
\usepackage{pifont}
\usepackage[colorlinks,
linkcolor=red,
anchorcolor=blue,
citecolor=blue,
urlcolor=magenta]{hyperref}
\begin{document}
\title{Learning 3D Gaussians for Extremely Sparse-View Cone-Beam CT Reconstruction}
\titlerunning{DIF-Gaussian}

\author{
Yiqun Lin\inst{1} \and 
Hualiang Wang\inst{1} \and 
Jixiang Chen\inst{1} \and 
Xiaomeng Li\inst{1,2}\thanks{Corresponding Author}
}
%index{Lin, Yiqun}
%index{Wang, Hualiang}
%index{Chen, Jixiang}
%index{Li, Xiaomeng}

\authorrunning{Y. Lin et al.}
\institute{
The Hong Kong University of Science and Technology \\
\email{eexmli@ust.hk} \and
HKUST Shenzhen-Hong Kong Collaborative Innovation Research \\ Institute, Futian, Shenzhen}

\maketitle              % typeset the header of the contribution

\newcommand{\nickname}{DIF-Gaussian}
\newcommand{\etal}{\textit{et~al.}}
\newcommand{\ie}{\textit{i.e.}}
\newcommand{\eg}{\textit{e.g.}}
\newcommand{\psp}{5pt}

\begin{abstract}

Cone-Beam Computed Tomography (CBCT) is an indispensable technique in medical imaging, yet the associated radiation exposure raises concerns in clinical practice. To mitigate these risks, sparse-view reconstruction has emerged as an essential research direction, aiming to reduce the radiation dose by utilizing fewer projections for CT reconstruction.
Although implicit neural representations have been introduced for sparse-view CBCT reconstruction, existing methods primarily focus on local 2D features queried from sparse projections, which is insufficient to process the more complicated anatomical structures, such as the chest.
To this end, we propose a novel reconstruction framework, namely \nickname{}, which leverages 3D Gaussians to represent the feature distribution in the 3D space, offering additional 3D spatial information to facilitate the estimation of attenuation coefficients. Furthermore, we incorporate test-time optimization during inference to further improve the generalization capability of the model. We evaluate \nickname{} on two public datasets, showing significantly superior reconstruction performance than previous state-of-the-art methods. The code is available at {\tt\small \url{https://github.com/xmed-lab/DIF-Gaussian}}.

\end{abstract}

\section{Introduction}

Computed Tomography (CT) is an indispensable technique in medical imaging, providing detailed internal views of the body to aid in diagnosis and treatment planning. 
Recently, Cone-Beam Computed Tomography (CBCT) has gained popularity due to its ability to offer high-resolution images with faster scanning speed~\cite{scarfe2006clinical} compared to traditional CT. 
Sparse-view reconstruction~\cite{ma2023freeseed,wu2022deep,wu2021drone,zha2022naf} has been introduced to reduce radiation exposure, where fewer projections are used without significantly reducing image quality. 
In this research, we follow \cite{lin2023learning} to study the problem of extremely sparse-view ($\leq$10) CBCT reconstruction, which is more challenging yet promising because extremely low radiation allows more frequent 3D scanning during the surgery, thereby enhancing surgical precision/adaptability, and meanwhile ensuring patient safety.

Sparse-view CBCT reconstruction aims to reconstruct 3D CT volumes from sparse 2D projections. Previously, FDK~\cite{feldkamp1984practical} was proposed based on filtered-backprojection (FBP) for CBCT reconstruction, while it requires hundreds of views to avoid streaking artifacts. Although ART-based methods~\cite{andersen1984simultaneous,gordon1970algebraic,pan2006variable} have been proposed for sparse-view reconstruction, their application is primarily effective in scenarios involving tens of views and the iterative optimization process is time-consuming. 
In recent years, learning-based methods become popular in sparse-view reconstruction with the development of deep learning technologies. 
Denoising methods~\cite{han2016deep,jin2017deep,ma2023freeseed,wu2022deep,zhang2018sparse} (2D$\rightarrow$2D) have been introduced for the reconstruction of conventional fan/parallel-beam CT. When adapted to CBCT through slice-wise processing, these methods struggle to ensure the spatial consistency of reconstructed 3D volumes. 
Voxel-based approaches~\cite{jiang2022mfct,kyung2023perspective,shen2019patient,ying2019x2ct} (2D$\rightarrow$3D) are proposed for single/orthogonal-view CBCT reconstruction, while extending these methods to sparse-view reconstruction encounters significant challenges due to the extremely high memory requirements, which ultimately lead to limited spatial resolution. 
Inspired by implicit neural representations~\cite{mildenhall2021nerf,ruckert2022neat}, researchers~\cite{lin2023learning,Lin_2024_CVPR,shen2022nerp,zha2022naf} represent CBCT as a continuous attenuation coefficient field, offering a new path for sparse-view CBCT reconstruction. 
Specifically, NAF~\cite{zha2022naf} and NeRP~\cite{shen2022nerp} are proposed to minimize the error between real and synthesized projections. However, per-sample optimization is time-consuming and unsuitable for extremely sparse-view reconstruction due to a lack of prior knowledge.
Lin~\etal~\cite{lin2023learning} propose DIF-Net trained on a CBCT dataset to learn an implicit mapping from extremely sparse projections to the intensity field. Nevertheless, only local semantic features are queried from 2D projections, which are insufficient for processing more complicated anatomical structures. 

\raggedbottom

3D Gaussians~\cite{kerbl20233d}, as a powerful and explicit representation of radiance fields, can be efficiently rendered by splatting~\cite{zwicker2001ewa}. Follow-up works extended 3D Gaussian Splatting~\cite{kerbl20233d} to downstream applications, such as mesh reconstruction~\cite{guedon2023sugar,li2024gaussianbody} and dynamic scene synthesis~\cite{luiten2023dynamic,yang2023neural,zhuendogs}, achieving state-of-the-art performance.
In this work, we propose a new reconstruction framework \nickname{} built on DIF-Net~\cite{lin2023learning} by leveraging 3D Gaussians to explicitly represent the feature distribution in the 3D space, which provides additional 3D spatial information to facilitate the estimation of attenuation coefficient values. 
A 3D Gaussian is defined by a collection of parameters: the 3D position, covariance matrix, and representative features. These parameters are derived from sparse-view projections and a predetermined set of points that indicate the initial positions of Gaussians. To be more specific, a 2D encoder first extracts sparse-view feature maps from the input projections. Subsequently, the initial position of each Gaussian serves as a reference point to query features from sparse-view features. Multi-layer perceptrons are then utilized to learn Gaussian parameters from queried features.
Therefore, hybrid features of points can be queried not only from sparse-view feature maps, but also from 3D Gaussians to enhance the representation.
To improve the generalization capability of our \nickname{}, we further propose test-time optimization (TTO) that can be applied during the model inference. Specifically, TTO fine-tunes the well-trained model with the test data (\ie, only sparse-view projections) based on the constraint derived from the foundational principles of X-ray imaging.
Finally, extensive experiments and ablative studies are conducted on two public datasets (chest and dental) with diverse anatomy, demonstrating the effectiveness and efficiency of our \nickname{} and TTO.

In summary, the contributions of our work mainly include 
1.) we are the first to introduce 3D Gaussians as an explicit feature representation in supervised CBCT reconstruction;
2.) we propose a new framework \nickname{}, where hybrid features of points are queried from learned 3D Gaussians and sparse-view projections to enhance the representation;
3.) we propose test-time optimization that can be applied during inference to further improve the generalization capability of \nickname{};
4.) experiments are conducted on two public datasets, showing that \nickname{} significantly outperforms previous methods by a remarkable margin.

% \newpage

\section{Methods}

In this section, we first describe the problem formulation of sparse-view CBCT reconstruction based on implicit neural representations. Then, we formally introduce the proposed \nickname{} framework and test-time optimization.

\subsection{Problem Formulation}

Following DIF-Net~\cite{lin2023learning}, we represent CT as a continuous field, where the model aims to learn an implicit mapping function $g$ such that $v = g(\mathcal{I}, \textbf{p})$, where $\mathcal{I} = \{I_1, \dots, I_K\}$ are $K$ sparse 2D projections, $\textbf{p} \in \mathbb{R}^3$ is an arbitrary point defined in the 3D space, and $v \in \mathbb{R}$ is the corresponding attenuation coefficient (or saying intensity in \cite{lin2023learning}) value. During training, projections are simulated from CT by digitally reconstructed radiographs (DRRs), and ground-truth attenuation coefficients are interpolated from the CT for point-wise supervision. In the inference stage, the model estimates the attenuation coefficient of the grid point centered at each CT voxel.

\subsection{\nickname{}: Learning 3D Gaussians}

\begin{figure}[t]
\centering 
\includegraphics[width=1.0\textwidth]{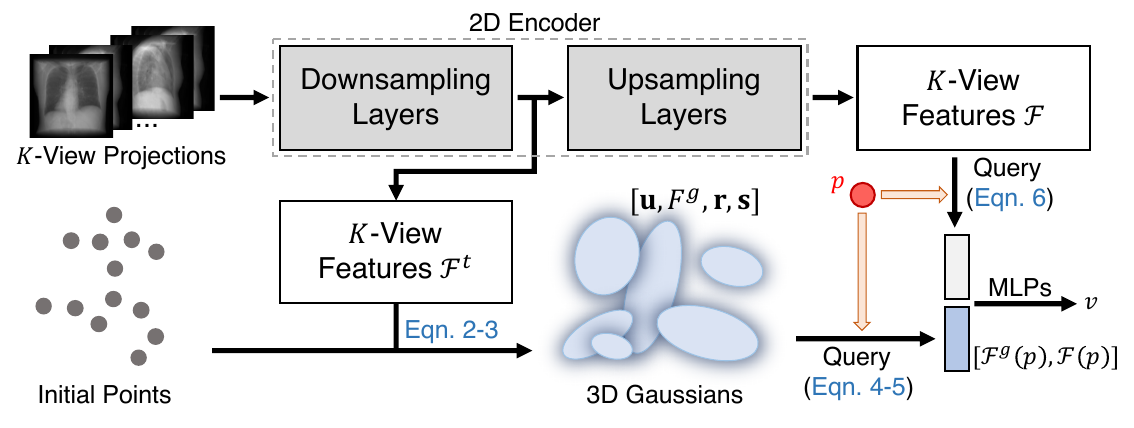}
\vspace{-8mm}
\caption{Overview of our \nickname{}. 3D Gaussian parameters are learned from $K$-view intermediate features $\mathcal{F}^t$. For a 3D point $p$, hybrid representative features are queried from Gaussians (3D) and $K$-view features (2D) to estimate its attenuation coefficient $v$.}
\label{fig:overview}
\end{figure}

Based on the above formulation, we develop a novel framework \nickname{} (see Figure~\ref{fig:overview}) for effective and efficient extremely sparse-view CBCT reconstruction. Overall, \nickname{} learns 3D Gaussians from sparse projections as an explicit 3D representation, and the features of a sampled point are queried from both sparse-view features and 3D Gaussians to enhance the representation.

\vspace{\psp{}}
\noindent
\textbf{3D Gaussians.} We define the properties of a 3D Gaussian -- 3D position $\textbf{u} \in \mathbb{R}^3$, covariance matrix $\Sigma \in \mathbb{R}^{3\times3}$, and representative features $F^g \in \mathbb{R}^{C^g}$. Inspired by \cite{kerbl20233d}, the anisotropic covariance matrix can be formulated as $\Sigma = L^TL$ and $L = M_rM_s \in \mathbb{R}^{3\times3}$, where $M_r, M_s \in \mathbb{R}^{3\times3}$ are rotation and scaling matrices. Additionally, $M_r, M_s$ can be determined by a 4-dimensional quaternion $\textbf{r} = [r_1, r_2, r_3, r_4] \in \mathbb{R}^4$ ($\|\textbf{r}\|_2=1$) and scaling factors $\textbf{s} = [s_1, s_2, s_3] \in \mathbb{R}^3$ defined in 3 dimensions, respectively. Specifically, $M_r$ and $M_s$ can be written as
\begin{equation}
M_r = 
\setlength\arraycolsep{4.5pt}
\begin{bmatrix}
    1 - 2r_3^2 - 2r_4^2 & 2r_2r_3 - 2r_1r_4 & 2r_2r_4 + 2r_1r_3 \\
    2r_2r_3 + 2r_1r_4 & 1 - 2r_2^2 - 2r_4^2 & 2r_3r_4 - 2r_1r_2 \\
    2r_2r_4 - 2r_1r_3 & 2r_3r_4 + 2r_1r_2 & 1 - 2r_2^2 - 2r_3^2 \\
\end{bmatrix}, \ 
M_s =
\setlength\arraycolsep{2pt}
\begin{bmatrix}
    s_1 & 0   & 0   \\
    0   & s_2 & 0   \\
    0   & 0   & s_3 \\
\end{bmatrix}.
\label{eq:gs_rs}
\end{equation}
Hence, a 3D Gaussian can be represented as a set of parameters $\{\textbf{u}, F^g, \textbf{r}, \textbf{s}\}$. 

\vspace{\psp{}}
\noindent
\textbf{Learn 3D Gaussians from Projections.} Given $K$-view projections, a shared 2D encoder is applied to extract semantic features. $K$-view feature maps ($t^\text{th}$ intermediate outputs of the 2D encoder) are denoted as $\mathcal{F}^t = \{F_1^t, \dots, F_K^t\} \subset \mathbb{R}^{W^t\times H^t\times {C^t}}$. For a 3D Gaussian with the predetermined initial position $\hat{\textbf{u}}$, we query view-specific features from $F_k^t$ using $\hat{\textbf{u}}$:
\begin{equation}
    F_k^t(\hat{\textbf{u}}) = \text{Interp}\big(F_k^t, \pi_k(\hat{\textbf{u}})\big) \in \mathbb{R}^{C^t},~\text{for}~k \in \{1, \dots, K\},
\end{equation}
where $\pi_k: \mathbb{R}^3 \rightarrow \mathbb{R}^2$ is the projection function of $k^\text{th}$ view, and $\text{Interp}(\cdot)$ indicates bilinear interpolation. $K$ queried features are aggregated with a max-pooling layer to obtain $\mathcal{F}^t(\hat{\textbf{u}}) = \text{Max-Pooling}(\{F_1^t(\hat{\textbf{u}}), \dots, F_K^t(\hat{\textbf{u}})\}) \in \mathbb{R}^C$. Multi-layer perceptions (MLPs) are then applied to learn Gaussian parameters:
\begin{equation}
    [\Delta\textbf{u}, F^g, \textbf{r}, \textbf{s}] = \text{MLPs}\big(\mathcal{F}^t(\hat{\textbf{u}})\big) \in \mathbb{R}^{3 + C^g + 4 + 3},
\label{eq:gs_params}
\end{equation}
where $\Delta\textbf{u}$ indicates the position offsets and the actual position of the Gaussian is $\textbf{u} = \hat{\textbf{u}} + \Delta\textbf{u}$. In practice, the initial position $\hat{\textbf{u}}$ is defined as the coordinate of a point and will not change after initialization. Hence, $N_g$ 3D Gaussians can be initialized with a set of points $\mathcal{P} = \{\hat{\textbf{u}}_1, \dots, \hat{\textbf{u}}_{N_g}\}$ indicating their initial positions and other parameters (\ie, $\{\Delta\textbf{u}, F^g, \textbf{r}, \textbf{s}\}$) can be then estimated from input projections based on initial positions (Eqn.~\ref{eq:gs_params}). In practice, we voxelize the space into $V$$\times$$V$$\times$$V$ voxels, and $\mathcal{P}$ are selected as the centroids of $V^3$ voxels.

\vspace{\psp{}}
\noindent
\textbf{Query Features from 3D Gaussians.} Given a point $\textbf{p} \in \mathbb{R}^3$ and a 3D Gaussian $\mathcal{G} = \{\textbf{u}, F^g, \textbf{r}, \textbf{s}\}$, we denote the covariance matrix of $\mathcal{G}$ as $\Sigma$, which is calculated using $\textbf{r}$ and $\textbf{s}$ (Eqn.~\ref{eq:gs_rs}). Then, the querying weight is defined as
\begin{equation}
    w(\textbf{p}, \mathcal{G}) = \frac{1}{\sqrt{(2\pi)^3|\Sigma|}}\cdot\text{exp}\Big(-\frac{1}{2}(\textbf{p} - \textbf{u})^T \Sigma^{-1} (\textbf{p} - \textbf{u})\Big).
\end{equation}
$N_g$ ($N_g = V^3$) 3D Gaussians are used to explicitly represent the feature distribution in the 3D space, which means that we can query features of $\textbf{p}$ directly from the 3D space based on these Gaussians:
\begin{equation}
\mathcal{F}^g(\textbf{p}) = \sum_{i=1}^{N_g} w(\textbf{p}, \mathcal{G}_i)\cdot F^g_i \in \mathbb{R}^{C^g},
\label{eq:Fgp}
\end{equation}
where $\mathcal{G}_i = \{\textbf{u}_i, F^g_i, \textbf{r}_i, \textbf{s}_i\}$ indicates $i^\text{th}$ Gaussian. Denoting $\mathcal{F} = \{F_1, \dots, F_K\} \subset \mathbb{R}^{W\times H\times C}$ as the $K$-view feature maps (final outputs of the 2D encoder), we can additionally query features of $\textbf{p}$ from $K$-view projections as 
\begin{equation}
\mathcal{F}(\textbf{p}) = \text{Max-Pooling}\Big(\big\{F_1(\textbf{p}), \dots, F_K(\textbf{p})\big\}\Big) \in \mathbb{R}^C,
\label{eq:Fp}
\end{equation}
where $F_k(\textbf{p}) = \text{Interp}\big(F_k, \pi_k(\textbf{p})\big)$ for $k \in \{1, \dots, K\}$. Then, features queried from 2D ($\mathcal{F}(\textbf{p})$ in Eqn.~\ref{eq:Fp}) and 3D ($\mathcal{F}^g(\textbf{p})$ in Eqn.~\ref{eq:Fgp}) are concatenated as the hybrid (2D+3D) representation of the point $\textbf{p}$. Finally, MLPs applied to estimate the corresponding attenuation coefficient $v = \text{MLPs}\big(\text{concat}\big[\mathcal{F}^g(\textbf{p}), \mathcal{F}(\textbf{p})\big]\big)$.

\vspace{\psp{}}
\noindent
\textbf{Implementation.} In practice, we follow \cite{lin2023learning} to use U-Net~\cite{ronneberger2015u} as the 2D encoder with the output channel $C = 128$. We choose the outputs of the final downsampling layer as $\mathcal{F}^t$ ($C^t = 1024$). Additionally, $C^g = 128$ and $V=12$ in our experiments. To simplify the calculation of Eqn.~\ref{eq:Fgp}, we choose the three nearest Gaussians of the point $\textbf{p}$ for approximation rather than using all Gaussians, where the distance is calculated based on the coordinates of $\textbf{p}$ and initial positions $\hat{\textbf{u}}$ of Gaussians. 10,000 points are randomly sampled from the 3D space for training, and point-wise mean-square-error (same as in \cite{lin2023learning}) is used for model optimization. Refer to the code (to be released later) for more details.

\subsection{Test-Time Optimization (TTO)}

Given a ray $R(\lambda) = \textbf{p}_s + \lambda(\textbf{p}_d - \textbf{p}_s)$ for $\lambda \in [0, 1]$, where $\textbf{p}_s$ is the X-ray source and $\textbf{p}_d$ is a point at the detector, the total energy attenuation accumulated by the ray (discrete approximation with $N_r + 1$ points) is given as
\begin{equation}
\begin{aligned}
e(R) 
\approx \big\|\textbf{p}_d - \textbf{p}_s\big\|_2\sum_{i=0}^{N_r} \mu\Big(\textbf{p}_s + \frac{i}{N_r}(\textbf{p}_d - \textbf{p}_s)\Big)\frac{1}{N_r},
\end{aligned}
\label{eq:imaging}
\end{equation}
where $\mu: \mathbb{R}^3 \rightarrow \mathbb{R}$ indicates the attenuation coefficient value of a given point. Numerically, the true $e(R)$ can be measured from the detector (at $\textbf{p}_d$), and $\mu$ should satisfy Eqn.~\ref{eq:imaging}.
Based on the above constraint, we further propose test-time optimization to improve the generalization capability of the well-train model $g$ during inference. Specifically, given sparse projections $\mathcal{I}$, the mapping function $\mu$ in Eqn.~\ref{eq:imaging} can be formulated as $\mu(\cdot) \equiv g(\mathcal{I}, \cdot)$. Then, we can optimize the projection error $\|e(R) - \hat{e}(R)\|_2$ to fine-tune $g$, where $e(R)$ is the true measurement of $\textbf{p}_d$ in the projection and $\hat{e}(R)$ is calculated using Eqn.~\ref{eq:imaging}.

% \newpage

\section{Experiments}

To validate the effectiveness of our proposed framework \nickname{} and test-time optimization (TTO), we compared the reconstruction performance with previous state-of-the-art (SoTA) methods on two publically available CT (or CBCT) datasets. Experiments demonstrate the superiority of our \nickname{} with a remarkable margin to SoTA, and our ablative study also shows that TTO can further improve the generalization capability of \nickname{} during the model inference.

\subsection{Experimental Settings}

\noindent
\textbf{Datasets.}
Experiments are conducted on two public datasets -- LUNA16~\cite{setio2017validation} and ToothFairy~\cite{cipriano2022deep}. LUNA16~\cite{setio2017validation} contains 888 chest CT scans, split into 738/50/ 100 for training/validation/testing; ToothFairy~\cite{cipriano2022deep} consists of 443 dental CBCT scans, split into 343/25/75 for training/validation/testing. We follow \cite{lin2023learning} to preprocess CT scans into 256$\times$256$\times$256 volumes with consistent spacing, \ie, [1.6, 1.6, 1.6]~mm for chest CT and [2.1, 5.4, 5.4]~mm for dental CBCT. The viewing angles of projections are uniformly sampled in the range of 180$^\circ$.

\vspace{\psp{}}
\noindent
\textbf{Training Details.} The proposed \nickname{} is implemented with PyTorch and trained on 2 $\sim$ 4 NVIDIA RTX 3090 GPUs (2 GPUs for 6/8-view and 4 GPUs for 10-view). The model is optimized using stochastic gradient descent (SGD) with a momentum of 0.98 and a learning rate of 0.01 (decayed per epoch by a factor of 0.001$^\text{1/MAX\_EPOCH}$). The model is trained for 400 epochs on LUNA16~\cite{setio2017validation} and 600 epochs on ToothFairy~\cite{cipriano2022deep} with a batch size of 8.

\vspace{\psp{}}
\noindent
\textbf{Evaluation Metrics.} Following previous works~\cite{lin2023learning,zha2022naf}, peak signal-to-noise ratio (PSNR) and structural similarity (SSIM) are evaluated to measure the reconstruction quality, where higher values indicate better performance.

\subsection{Results}

\begin{figure}[t]
\centering 
\includegraphics[width=1.0\textwidth]{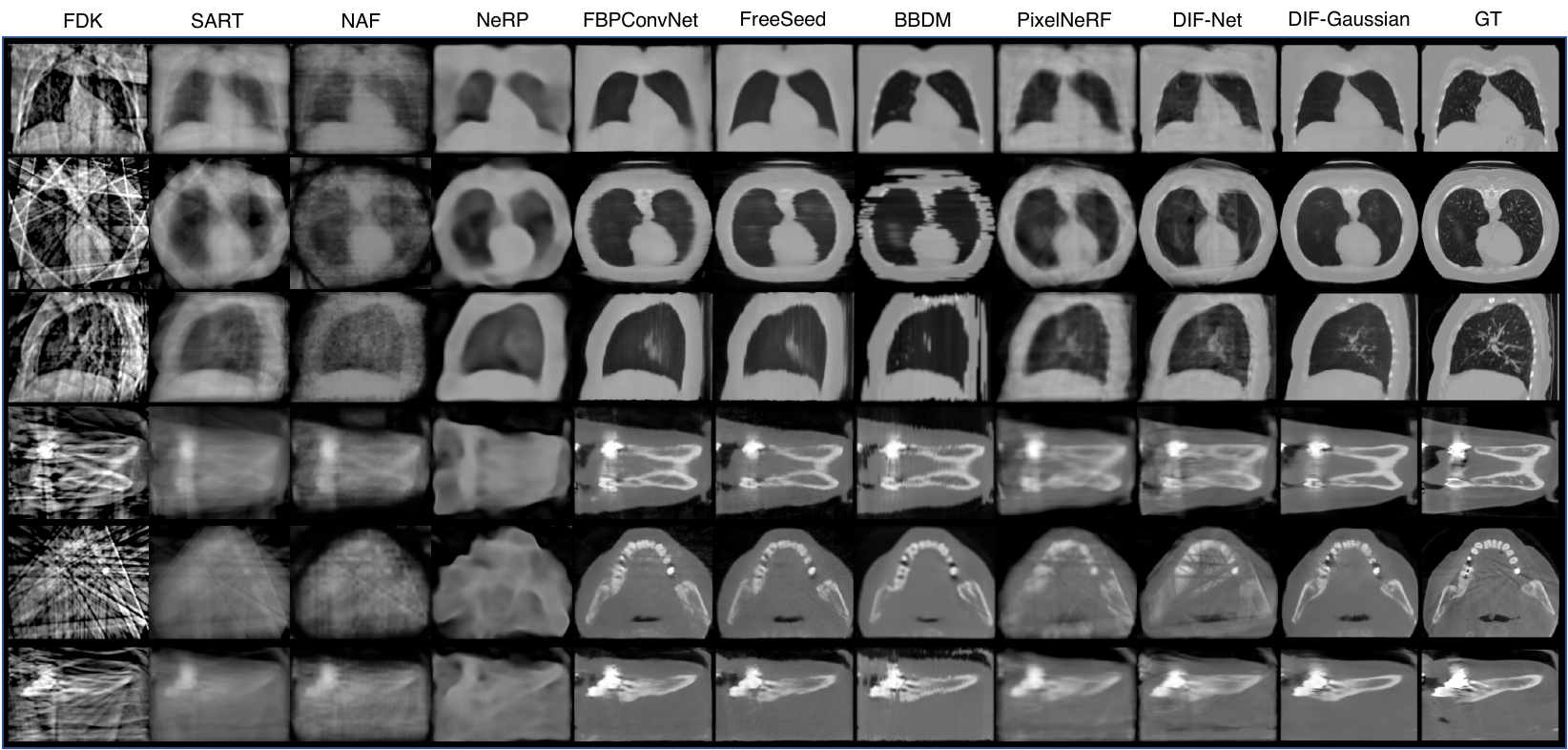}
\vspace{-7mm}
\caption{Visualization of different methods. Setting: 6-view reconstruction (resolution = 256$^\text{3}$). Top/bottom 3 rows: LUNA16/ToothFairy axial, coronal, and sagittal slices.}
\label{fig:vis}
\end{figure}

\noindent
\textbf{Comparison with SoTA.} 
In Table~\ref{tab:all}, we compare our proposed \nickname{} with self-supervised methods, including FDK~\cite{feldkamp1984practical}, SART~\cite{andersen1984simultaneous}, NAF~\cite{zha2022naf}, and NeRP~\cite{shen2022nerp}, where no training data is required, and the optimization is conducted only based on sparse projections. Additionally, we compare data-driven methods, including denoising-based (FBPConvNet~\cite{jin2017deep}, FreeSeed~\cite{ma2023freeseed}, and BBDM~\cite{li2023bbdm}) and implicit neural representation (INR)-based (PixelNeRF~\cite{yu2021pixelnerf} and DIF-Net~\cite{lin2023learning}). Experiments are conducted with different numbers (6/8/10) of projection views, and the reconstruction resolution is 256$\times$256$\times$256. Quantitative and qualitative results are shown in Table~\ref{tab:all} and Figure~\ref{fig:vis}, respectively. The quality of CT reconstructed by self-supervised methods is very poor as no prior knowledge is given, and the number of views is extremely limited. Denoising-based methods often suffer from jitter near organ boundaries because slice-wise (2D) denoising cannot guarantee 3D spatial consistency. Although previous INR-based methods can reconstruct CT with satisfactory contours, details are severely lost as the anatomical structures of the chest and dental CT are more complicated than the knee~\cite{lin2023learning}. Our \nickname{} significantly outperforms all compared methods on both two datasets by a remarkable margin. Furthermore, it is worth noting that even with only 6 views, our proposed \nickname{} can still reconstruct CT with better image quality than other methods with 10 views.

\begin{table}[t]
\centering
\caption{Quantitaive evaluation of compared methods on two public datasets with different numbers of projection views (6/8/10). The reconstruction resolution is 256$^\text{3}$. PSNR (dB) and SSIM (10$^\text{-2}$) are evaluated to measure the reconstruction quality (higher is better). Best values are \textbf{bolded}, and the second-best values are \underline{underlined}.}
\vspace{-6pt}
\label{tab:all}
\setlength{\tabcolsep}{2pt}
\resizebox{1.0\linewidth}{!}{
\begin{tabular}{l|c|ccc|ccc}
\toprule[1.2pt]
\multirow{2}{*}{Method} & \multirow{2}{*}{Type} & \multicolumn{3}{c|}{LUNA16~\cite{setio2017validation} (Chest CT)} & \multicolumn{3}{c}{ToothFairy~\cite{cipriano2022deep} (Dental CBCT)} \\ \cline{3-8}
 &  & 6-View & 8-View & 10-View & 6-View & 8-View & 10-View \\ 
 \hline
FDK~\cite{feldkamp1984practical} & \multirow{4}{*}{\begin{tabular}[c]{@{}c@{}}Self-\\ Supervised\end{tabular}} & 15.34$|$35.78 & 16.58$|$37.89 & 17.40$|$39.85 & 17.07$|$39.90 & 18.42$|$43.29 & 19.58$|$47.21 \\
SART~\cite{andersen1984simultaneous} &  & 19.70$|$64.36 & 20.06$|$67.80 & 20.23$|$70.23 & 20.04$|$64.98 & 21.92$|$67.86 & 22.82$|$71.53 \\
NAF~\cite{zha2022naf} & & 18.76$|$54.16 & 20.51$|$60.84 & 22.17$|$62.22 & 20.58$|$63.52 & 22.39$|$67.24 & 23.84$|$72.52 \\
NeRP~\cite{shen2022nerp} & & 23.55$|$74.46 & 25.83$|$80.67 & 26.12$|$81.30 & 21.77$|$72.06 & 24.18$|$78.83 & 25.99$|$82.08 \\ 
\hline
FBPConvNet~\cite{jin2017deep} & \multirow{3}{*}{\begin{tabular}[c]{@{}c@{}}Data-Driven:\\Denoising\end{tabular}} & 24.38$|$77.57 & 24.87$|$78.86 & 25.90$|$80.03 & \underline{27.22}$|$79.33 & \underline{27.72}$|$81.90 & \underline{28.13}$|$83.51 \\
FreeSeed~\cite{ma2023freeseed} & & \underline{25.59}$|$77.36 & \underline{26.86}$|$78.92 & \underline{27.23}$|$79.25 & 26.35$|$78.98 & 27.08$|$81.38 & 27.63$|$84.40 \\
BBDM~\cite{li2023bbdm} & & 24.78$|$77.03 & 25.81$|$78.06 & 26.35$|$79.38 & 26.29$|$78.57 & 27.28$|$80.33 & 28.00$|$83.96 \\
\hline
PixelNeRF~\cite{yu2021pixelnerf} & \multirow{3}{*}{\begin{tabular}[c]{@{}c@{}}Data-Driven:\\ INR-based\end{tabular}} & 24.66$|$78.68 & 25.04$|$80.57 & 25.39$|$82.13 & 24.85$|$80.91 & 25.37$|$82.11 & 25.90$|$83.25 \\
DIF-Net~\cite{lin2023learning} & & 25.55$|$\underline{84.40} & 26.09$|$\underline{85.07} & 26.67$|$\underline{86.09} & 25.78$|$\underline{83.62} & 26.29$|$\underline{84.81} & 26.90$|$\underline{86.42} \\
\nickname{} (\textit{ours}) & & \textbf{28.48}$|$\textbf{91.31} & \textbf{29.46}$|$\textbf{92.57} & \textbf{30.01}$|$\textbf{93.29} & \textbf{27.92}$|$\textbf{90.19} & \textbf{28.35}$|$\textbf{90.76} & \textbf{29.24}$|$\textbf{92.13} \\
\bottomrule[1.2pt]
\end{tabular}
}
\end{table}

\begin{table}[t]
\centering
\begin{minipage}[t]{.56\linewidth}
\caption{Comparison regarding the number of parameters, time and memory (MB) for the model training and inference. Setting: 6-view chest reconstruction (resolution = 256$^\text{3}$). Batch size is set to 1 for training memory calculation.}
\vspace{-6pt}
\label{tab:time}
\setlength{\tabcolsep}{2pt}
\resizebox{1.0\textwidth}{!}{
\begin{tabular}{l|r|rr|rr}
\toprule[1.2pt]
\multirow{2}{*}{Method} & \multirow{2}{*}{\begin{tabular}[c]{@{}c@{}}Param.\\ (M)\end{tabular}} & \multicolumn{2}{c|}{Training} & \multicolumn{2}{c}{Inference} \\ \cline{3-6} 
 &  & Time (h) & ~~Mem. & Time (s) & ~~Mem. \\ \hline
FDK~\cite{feldkamp1984practical} 
& \multicolumn{1}{c|}{-} & \multicolumn{1}{c}{-} & \multicolumn{1}{c|}{-} & 0.3~ & \multicolumn{1}{c}{-} \\
SART~\cite{andersen1984simultaneous} & \multicolumn{1}{c|}{-} & \multicolumn{1}{c}{-} & \multicolumn{1}{c|}{-} & 60.2~ & 327~ \\
NAF~\cite{zha2022naf} & 14.3~ & \multicolumn{1}{c}{-} & \multicolumn{1}{c|}{-} & 433.1~ & 2933~ \\
NeRP~\cite{shen2022nerp} & 0.7~ & \multicolumn{1}{c}{-} & \multicolumn{1}{c|}{-} & 937.5~ & 8229~ \\ \hline
FBPConvNet~\cite{jin2017deep} & 34.6~ & 2.6~~~~ & 2821~ & 3.7~ & 2169~ \\
FreeSeed~\cite{ma2023freeseed} & 8.7~ & 2.2~~~~ & 2197~ & 1.7~ & 1931~ \\
BBDM~\cite{li2023bbdm} & 237.1~ & 11.7~~~~ & 10345~ & 2176.5~ & 6481~ \\ \hline
PixelNeRF~\cite{yu2021pixelnerf} & 24.7~ & 10.3~~~~ & 4963~ & 40.4~ & 9693~ \\
DIF-Net~\cite{lin2023learning} & 31.1~ & 4.9~~~~ & 5447~ & 1.1~ & 4409~ \\
\nickname{} (\textit{ours}) & 31.7~ & 5.3~~~~ & 5957~ & 1.8~ & 5031~ \\
\bottomrule[1.2pt]
\end{tabular}
}
\end{minipage}
\hfill
\begin{minipage}[t]{.41\linewidth}
\centering
\caption{Ablation on test-time optimization (TTO) and the number ($N_g$) of Gaussians. PSNR (dB) and SSIM (10$^\text{-2}$) are evaluated. Experiments are conducted on 6-view reconstruction (resolution = 256$^\text{3}$).}
\vspace{-6pt}
\label{tab:tto}
\setlength{\tabcolsep}{7pt}
\resizebox{1.0\textwidth}{!}{
\begin{tabular}{l|c|c|c}
\toprule[1.2pt]
\multirow{2}{*}{Training Set} & \multirow{2}{*}{TTO} & \multicolumn{2}{c}{Test Set} \\ \cline{3-4}
 & & LUNA16 & ToothFairy \\ \hline
\multirow{2}{*}{LUNA16} 
 & \ding{55} & 28.48$|$91.31 & \multirow{2}{*}{-} \\
 & \ding{51} & 28.59$|$91.52 & \\ \hline
\multirow{2}{*}{ToothFairy} 
 & \ding{55} & \multirow{2}{*}{-} & 27.92$|$90.19 \\
 & \ding{51} &  & 28.04$|$90.38 \\ \hline
\multirow{2}{*}{\begin{tabular}[l]{@{}l@{}}LUNA16\\+ToothFairy\end{tabular}} 
 & \ding{55} & 27.14$|$89.40 & 26.92$|$88.37 \\
 & \ding{51} & 27.44$|$90.62 & 27.23$|$89.29 \\
\bottomrule[1.2pt]
\end{tabular}
}
\setlength{\tabcolsep}{2pt}
\resizebox{1.0\textwidth}{!}{
\begin{tabular}{l|c|c|c}
& $N_g = 8^3$ & $N_g = 12^3$ & $N_g = 16^3$ \\ \hline
LUNA16 & 28.42$|$91.18 & 28.48$|$91.31 & 28.48$|$91.32 \\ \hline
ToothFairy & 27.82$|$90.01 & 27.92$|$90.19 & 27.93$|$90.19 \\
\bottomrule[1.2pt]
\end{tabular}
}
\end{minipage}
\end{table}

\vspace{\psp{}}
\noindent
\textbf{Efficiency Analysis.}
In Table~\ref{tab:time}, we compare the training and inference efficiency of different reconstruction methods. For self-supervised methods, the inference includes per-sample optimization and network inference. Self-supervised methods (except FDK~\cite{feldkamp1984practical}) often require a long time for optimization during the reconstruction. BBDM~\cite{li2023bbdm} reconstructs CT with the lowest speed due to the complex, iterative nature of diffusion models, requiring many sequential steps to refine. More importantly, our \nickname{} significantly improves the reconstruction performance, yet maintaining reconstruction efficiency that is comparable to prior DIF-Net~\cite{lin2023learning}. Note that TTO is not incorporated into \nickname{} and will be discussed separately in the ablation study (next paragraph).

\vspace{\psp{}}
\noindent
\textbf{Ablation Study.} In Table~\ref{tab:tto}, we compare the performance of inference with and without test-time optimization (TTO) in different experimental settings. Results show that TTO can improve the reconstruction performance in both two datasets. Specifically, the extent of improvement depends on how closely the test data aligns with the overall distribution of training data. For instance, the improvement is 0.1/0.2 PSNR/SSIM for a model trained and tested on LUNA16, whereas the improvement is more substantial (0.3/0.8 PSNR/SSIM) for a model trained on LUNA16+ToothFairy and tested on LUNA16. Additionally, we compare different numbers ($N_g = 8^3/12^3/16^3$) of Gaussians used in \nickname{} and find that $N_g = 12^3$ is the optimal choice in both two datasets for balancing performance improvement and processing efficiency.

\section{Conclusion}

In this study, we present a new framework \nickname{} for extremely sparse-view CBCT reconstruction. Instead of solely relying on features queried from 2D sparse-view projections (like DIF-Net~\cite{lin2023learning}), 3D Gaussians are introduced to provide additional 3D spatial information and facilitate the learning of attenuation coefficients. 
Additionally, test-time optimization (TTO) is proposed to further improve the generalization capability of the model during inference. Experiments conducted on two public datasets (chest CT and dental CBCT) demonstrate the superior reconstruction performance of our \nickname{}, as well as the effectiveness of TTO. 
In our experiments, the predetermined initial position of a Gaussian is at the centroid of a voxel. Alternatively, the initial positions could be points located on the boundary of an organ or uniformly distributed within specific organs. However, exploring these alternatives involves additional tasks (\eg, boundary/organ detection), which will be left as our future work.

\vspace{6pt}
\noindent
\textbf{Acknowledgements.} 
This work was supported in part by grants from the National Natural Science Foundation of China under Grant No. 62306254, grants from the Foshan HKUST Projects under Grants FSUST21-HKUST10E and FSUST21-HKUST11E and Project of Hetao Shenzhen-Hong Kong Science and Technology Innovation Cooperation Zone (HZQB-KCZYB-2020083).

\vspace{6pt}
\noindent
\textbf{Disclosure of Interests.}
The authors have no competing interests to declare that are relevant to the content of this article.

% \newpage
\bibliographystyle{splncs04}
\bibliography{Paper-0250}

\begin{thebibliography}{10}
\providecommand{\url}[1]{\texttt{#1}}
\providecommand{\urlprefix}{URL }
\providecommand{\doi}[1]{https://doi.org/#1}

\bibitem{andersen1984simultaneous}
Andersen, A.H., Kak, A.C.: Simultaneous algebraic reconstruction technique (sart): a superior implementation of the art algorithm. Ultrasonic imaging  \textbf{6}(1),  81--94 (1984)

\bibitem{cipriano2022deep}
Cipriano, M., Allegretti, S., Bolelli, F., Di~Bartolomeo, M., Pollastri, F., Pellacani, A., Minafra, P., Anesi, A., Grana, C.: Deep segmentation of the mandibular canal: a new 3d annotated dataset of cbct volumes. IEEE Access  \textbf{10},  11500--11510 (2022)

\bibitem{feldkamp1984practical}
Feldkamp, L.A., Davis, L.C., Kress, J.W.: Practical cone-beam algorithm. Josa a  \textbf{1}(6),  612--619 (1984)

\bibitem{gordon1970algebraic}
Gordon, R., Bender, R., Herman, G.T.: Algebraic reconstruction techniques (art) for three-dimensional electron microscopy and x-ray photography. Journal of theoretical Biology  \textbf{29}(3),  471--481 (1970)

\bibitem{guedon2023sugar}
Gu{\'e}don, A., Lepetit, V.: Sugar: Surface-aligned gaussian splatting for efficient 3d mesh reconstruction and high-quality mesh rendering. arXiv preprint arXiv:2311.12775  (2023)

\bibitem{han2016deep}
Han, Y.S., Yoo, J., Ye, J.C.: Deep residual learning for compressed sensing ct reconstruction via persistent homology analysis. arXiv preprint arXiv:1611.06391  (2016)

\bibitem{jiang2022mfct}
Jiang, Y.: Mfct-gan: multi-information network to reconstruct ct volumes for security screening. Journal of Intelligent Manufacturing and Special Equipment  (2022)

\bibitem{jin2017deep}
Jin, K.H., McCann, M.T., Froustey, E., Unser, M.: Deep convolutional neural network for inverse problems in imaging. IEEE Transactions on Image Processing  \textbf{26}(9),  4509--4522 (2017)

\bibitem{kerbl20233d}
Kerbl, B., Kopanas, G., Leimk{\"u}hler, T., Drettakis, G.: 3d gaussian splatting for real-time radiance field rendering. ACM Transactions on Graphics  \textbf{42}(4) (2023)

\bibitem{kyung2023perspective}
Kyung, D., Jo, K., Choo, J., Lee, J., Choi, E.: Perspective projection-based 3d ct reconstruction from biplanar x-rays. In: ICASSP 2023-2023 IEEE International Conference on Acoustics, Speech and Signal Processing (ICASSP). pp.~1--5. IEEE (2023)

\bibitem{li2023bbdm}
Li, B., Xue, K., Liu, B., Lai, Y.K.: Bbdm: Image-to-image translation with brownian bridge diffusion models. In: Proceedings of the IEEE/CVF Conference on Computer Vision and Pattern Recognition. pp. 1952--1961 (2023)

\bibitem{li2024gaussianbody}
Li, M., Yao, S., Xie, Z., Chen, K., Jiang, Y.G.: Gaussianbody: Clothed human reconstruction via 3d gaussian splatting. arXiv preprint arXiv:2401.09720  (2024)

\bibitem{lin2023learning}
Lin, Y., Luo, Z., Zhao, W., Li, X.: Learning deep intensity field for extremely sparse-view cbct reconstruction. In: Medical Image Computing and Computer Assisted Intervention -- MICCAI 2023. pp. 13--23. Springer Nature Switzerland, Cham (2023)

\bibitem{Lin_2024_CVPR}
Lin, Y., Yang, J., Wang, H., Ding, X., Zhao, W., Li, X.: C{\textasciicircum}2rv: Cross-regional and cross-view learning for sparse-view cbct reconstruction. In: Proceedings of the IEEE/CVF Conference on Computer Vision and Pattern Recognition (CVPR). pp. 11205--11214 (June 2024)

\bibitem{luiten2023dynamic}
Luiten, J., Kopanas, G., Leibe, B., Ramanan, D.: Dynamic 3d gaussians: Tracking by persistent dynamic view synthesis. arXiv preprint arXiv:2308.09713  (2023)

\bibitem{ma2023freeseed}
Ma, C., Li, Z., Zhang, J., Zhang, Y., Shan, H.: Freeseed: Frequency-band-aware and self-guided network for sparse-view ct reconstruction. In: International Conference on Medical Image Computing and Computer-Assisted Intervention. pp. 250--259. Springer (2023)

\bibitem{mildenhall2021nerf}
Mildenhall, B., Srinivasan, P.P., Tancik, M., Barron, J.T., Ramamoorthi, R., Ng, R.: Nerf: Representing scenes as neural radiance fields for view synthesis. Communications of the ACM  \textbf{65}(1),  99--106 (2021)

\bibitem{pan2006variable}
Pan, J., Zhou, T., Han, Y., Jiang, M.: Variable weighted ordered subset image reconstruction algorithm. International Journal of Biomedical Imaging  \textbf{2006} (2006)

\bibitem{ronneberger2015u}
Ronneberger, O., Fischer, P., Brox, T.: U-net: Convolutional networks for biomedical image segmentation. In: Medical Image Computing and Computer-Assisted Intervention--MICCAI 2015: 18th International Conference, Munich, Germany, October 5-9, 2015, Proceedings, Part III 18. pp. 234--241. Springer (2015)

\bibitem{ruckert2022neat}
R{\"u}ckert, D., Wang, Y., Li, R., Idoughi, R., Heidrich, W.: Neat: Neural adaptive tomography. ACM Transactions on Graphics (TOG)  \textbf{41}(4),  1--13 (2022)

\bibitem{scarfe2006clinical}
Scarfe, W.C., Farman, A.G., Sukovic, P., et~al.: Clinical applications of cone-beam computed tomography in dental practice. Journal-Canadian Dental Association  \textbf{72}(1), ~75 (2006)

\bibitem{setio2017validation}
Setio, A.A.A., Traverso, A., De~Bel, T., Berens, M.S., Van Den~Bogaard, C., Cerello, P., Chen, H., Dou, Q., Fantacci, M.E., Geurts, B., et~al.: Validation, comparison, and combination of algorithms for automatic detection of pulmonary nodules in computed tomography images: the luna16 challenge. Medical image analysis  \textbf{42},  1--13 (2017)

\bibitem{shen2022nerp}
Shen, L., Pauly, J., Xing, L.: Nerp: implicit neural representation learning with prior embedding for sparsely sampled image reconstruction. IEEE Transactions on Neural Networks and Learning Systems  (2022)

\bibitem{shen2019patient}
Shen, L., Zhao, W., Xing, L.: Patient-specific reconstruction of volumetric computed tomography images from a single projection view via deep learning. Nature biomedical engineering  \textbf{3}(11),  880--888 (2019)

\bibitem{wu2022deep}
Wu, W., Guo, X., Chen, Y., Wang, S., Chen, J.: Deep embedding-attention-refinement for sparse-view ct reconstruction. IEEE Transactions on Instrumentation and Measurement  (2022)

\bibitem{wu2021drone}
Wu, W., Hu, D., Niu, C., Yu, H., Vardhanabhuti, V., Wang, G.: Drone: Dual-domain residual-based optimization network for sparse-view ct reconstruction. IEEE Transactions on Medical Imaging  \textbf{40}(11),  3002--3014 (2021)

\bibitem{yang2023neural}
Yang, C., Wang, K., Wang, Y., Yang, X., Shen, W.: Neural lerplane representations for fast 4d reconstruction of deformable tissues. arXiv preprint arXiv:2305.19906  (2023)

\bibitem{ying2019x2ct}
Ying, X., Guo, H., Ma, K., Wu, J., Weng, Z., Zheng, Y.: X2ct-gan: reconstructing ct from biplanar x-rays with generative adversarial networks. In: Proceedings of the IEEE/CVF conference on computer vision and pattern recognition. pp. 10619--10628 (2019)

\bibitem{yu2021pixelnerf}
Yu, A., Ye, V., Tancik, M., Kanazawa, A.: pixelnerf: Neural radiance fields from one or few images. In: Proceedings of the IEEE/CVF Conference on Computer Vision and Pattern Recognition. pp. 4578--4587 (2021)

\bibitem{zha2022naf}
Zha, R., Zhang, Y., Li, H.: Naf: Neural attenuation fields for sparse-view cbct reconstruction. In: Medical Image Computing and Computer Assisted Intervention--MICCAI 2022: 25th International Conference, Singapore, September 18--22, 2022, Proceedings, Part VI. pp. 442--452. Springer (2022)

\bibitem{zhang2018sparse}
Zhang, Z., Liang, X., Dong, X., Xie, Y., Cao, G.: A sparse-view ct reconstruction method based on combination of densenet and deconvolution. IEEE transactions on medical imaging  \textbf{37}(6),  1407--1417 (2018)

\bibitem{zhuendogs}
Zhu, L., Wang, Z., Jin, Z., Lin, G., Yu, L.: Deformable endoscopic tissues reconstruction with gaussian splatting. arXiv preprint arXiv:2401.11535  (2024)

\bibitem{zwicker2001ewa}
Zwicker, M., Pfister, H., Van~Baar, J., Gross, M.: Ewa volume splatting. In: Proceedings Visualization, 2001. VIS'01. pp. 29--538. IEEE (2001)

\end{thebibliography}

\end{document}